\PassOptionsToPackage{unicode}{hyperref}
\PassOptionsToPackage{hyphens}{url}
\PassOptionsToPackage{dvipsnames,svgnames,x11names}{xcolor}
\documentclass[
  12pt]{article}

\usepackage{amsmath,amssymb}
\usepackage{iftex}
\ifPDFTeX
  \usepackage[T1]{fontenc}
  \usepackage[utf8]{inputenc}
  \usepackage{textcomp} 
\else 
  \usepackage{unicode-math}
  \defaultfontfeatures{Scale=MatchLowercase}
  \defaultfontfeatures[\rmfamily]{Ligatures=TeX,Scale=1}
\fi
\usepackage{lmodern}
\ifPDFTeX\else  
\fi
\IfFileExists{upquote.sty}{\usepackage{upquote}}{}
\IfFileExists{microtype.sty}{
  \usepackage[]{microtype}
  \UseMicrotypeSet[protrusion]{basicmath} 
}{}
\makeatletter
\@ifundefined{KOMAClassName}{
  \IfFileExists{parskip.sty}{%
    \usepackage{parskip}
  }{
    \setlength{\parindent}{0pt}
    \setlength{\parskip}{6pt plus 2pt minus 1pt}}
}{
  \KOMAoptions{parskip=half}}
\makeatother
\usepackage{xcolor}
\ifx\paragraph\undefined\else
  \let\oldparagraph\paragraph
  \renewcommand{\paragraph}[1]{\oldparagraph{#1}\mbox{}}
\fi
\ifx\subparagraph\undefined\else
  \let\oldsubparagraph\subparagraph
  \renewcommand{\subparagraph}[1]{\oldsubparagraph{#1}\mbox{}}
\fi

\usepackage{longtable,booktabs,array}
\usepackage{calc} 
\usepackage{etoolbox}
\makeatletter
\patchcmd\longtable{\par}{\if@noskipsec\mbox{}\fi\par}{}{}
\makeatother
\IfFileExists{footnotehyper.sty}{\usepackage{footnotehyper}}{\usepackage{footnote}}
\makesavenoteenv{longtable}
\usepackage{graphicx}
\makeatletter
\def\maxwidth{\ifdim\Gin@nat@width>\linewidth\linewidth\else\Gin@nat@width\fi}
\def\maxheight{\ifdim\Gin@nat@height>\textheight\textheight\else\Gin@nat@height\fi}
\makeatother
\setkeys{Gin}{width=\maxwidth,height=\maxheight,keepaspectratio}
\makeatletter
\def\fps@figure{htbp}
\makeatother

\usepackage{booktabs}
\usepackage{longtable}
\usepackage{array}
\usepackage{multirow}
\usepackage{wrapfig}
\usepackage{float}
\usepackage{colortbl}
\usepackage{pdflscape}
\usepackage{tabu}
\usepackage{threeparttable}
\usepackage{threeparttablex}
\usepackage[normalem]{ulem}
\usepackage{makecell}
\usepackage{xcolor}
\usepackage{xcolor}
\usepackage{setspace}
\usepackage{fontawesome5}
\definecolor{mygreen}{HTML}{00b100}
\newcommand*\samethanks[1][\value{footnote}]{\footnotemark[#1]}
\makeatletter
\@ifpackageloaded{caption}{}{\usepackage{caption}}
\AtBeginDocument{%
\ifdefined\contentsname
  \renewcommand*\contentsname{Table of contents}
\else
  \newcommand\contentsname{Table of contents}
\fi
\ifdefined\listfigurename
  \renewcommand*\listfigurename{List of Figures}
\else
  \newcommand\listfigurename{List of Figures}
\fi
\ifdefined\listtablename
  \renewcommand*\listtablename{List of Tables}
\else
  \newcommand\listtablename{List of Tables}
\fi
\ifdefined\figurename
  \renewcommand*\figurename{Figure}
\else
  \newcommand\figurename{Figure}
\fi
\ifdefined\tablename
  \renewcommand*\tablename{Table}
\else
  \newcommand\tablename{Table}
\fi
}
\@ifpackageloaded{float}{}{\usepackage{float}}
\floatstyle{ruled}
\@ifundefined{c@chapter}{\newfloat{codelisting}{h}{lop}}{\newfloat{codelisting}{h}{lop}[chapter]}
\floatname{codelisting}{Listing}

\makeatother
\makeatletter
\@ifpackageloaded{caption}{}{\usepackage{caption}}
\@ifpackageloaded{subcaption}{}{\usepackage{subcaption}}
\makeatother
\makeatletter
\@ifpackageloaded{tcolorbox}{}{\usepackage[skins,breakable]{tcolorbox}}
\makeatother
\makeatletter
\@ifundefined{shadecolor}{\definecolor{shadecolor}{rgb}{.97, .97, .97}}
\makeatother
\ifLuaTeX
  \usepackage{selnolig}  
\fi
\usepackage[]{natbib}
\IfFileExists{bookmark.sty}{\usepackage{bookmark}}{\usepackage{hyperref}}
\IfFileExists{xurl.sty}{\usepackage{xurl}}{} 
\hypersetup{
  pdftitle={Automated grading workflows for providing personalized feedback to open-ended data science assignments},
  pdfauthor={Federica Zoe Ricci; Catalina Mari Medina; Mine Dogucu},
  pdfkeywords={data science education, statistics
education, R, formative assessment, fair grading, reproducible
teaching},
  colorlinks=true,
  linkcolor={blue},
  filecolor={Maroon},
  citecolor={Blue},
  urlcolor={Blue},
  pdfcreator={LaTeX via pandoc}}

\begin{document}

\def\spacingset#1{\renewcommand{\baselinestretch}%
{#1}\small\normalsize} \spacingset{1}


\date{February 29, 2024}
\title{\bf Automated grading workflows for providing personalized
feedback to open-ended data science assignments}
\author{
Federica Zoe Ricci\thanks{Gratefully acknowledges funding by the
Hasso-Plattner-Institute Research Center in Machine Learning and Data
Science at UCI.}\\
and\\Catalina Mari Medina\thanks{Gratefully acknowledge funding by NSF
IIS award 2123366.}\\
and\\Mine Dogucu\samethanks\\
Department of Statistics, University of California Irvine\\
}
\maketitle

\bigskip
\bigskip
\begin{abstract}
Open-ended assignments - such as lab reports and semester-long projects
- provide data science and statistics students with opportunities for
developing communication, critical thinking, and creativity skills.
However, providing grades and formative feedback to open-ended
assignments can be very time consuming and difficult to do consistently
across students. In this paper, we discuss the steps of a typical
grading workflow and highlight which steps can be automated in an
approach that we call automated grading workflow. We illustrate how
gradetools, a new R package, implements this approach within RStudio to
facilitate efficient and consistent grading while providing
individualized feedback. By outlining the motivations behind the
development of this package and the considerations underlying its
design, we hope this article will provide data science and statistics
educators with ideas for improving their grading workflows, possibly
developing new grading tools or considering use gradetools as their
grading workflow assistant.
\end{abstract}

\noindent%
{\it Keywords:} data science education, statistics
education, R, formative assessment, fair grading, reproducible teaching
\vfill

\newpage
\spacingset{1.9} 
\ifdefined\Shaded\renewenvironment{Shaded}{\begin{tcolorbox}[breakable, frame hidden, boxrule=0pt, borderline west={3pt}{0pt}{shadecolor}, interior hidden, sharp corners, enhanced]}{\end{tcolorbox}}\fi

\hypertarget{sec-intro}{%
\section{Introduction}\label{sec-intro}}

From a learner's standpoint, assessments are fundamental moments of the
learning process. Assessments (especially \emph{formative} ones
\citep{dixson2016formative}) give students opportunities to practice,
interiorize, deepen and demonstrate the concepts learned with lectures
and readings. By requiring the use of class material for answering
questions and solving problems, assessments reveal what parts of the
syllabus are well understood and what parts need, instead, revision.
This is crucial for both students and instructors to make informed
decisions on how to improve, respectively, their learning and teaching
strategies \citep{pai}.

Following up on assessment activities, teachers can give students
qualitative (e.g., ``The source of the data utilized in your analysis
was not specified.'') and quantitative (e.g., ``9 out of 10'') comments.
To distinguish the former from the latter, in this paper we will reserve
the term \emph{feedback} for qualitative comments and we will call
\emph{grades} the scores that quantitatively summarize how well a
student did on the assignment. Grades are a convenient summary of
students' performance \citep{lipnevich2009really}. On the other hand, as
stated by the Guidelines for Assessment and Instruction in Statistics
Education \citep{GAISE2016}, assessment must receive feedback in order
to lead to learning, and teachers should provide assessment \emph{and}
feedback throughout their courses.

The importance of feedback is underlined by several studies in the
literature on education. Reviewing research on formative assessment,
Black and William \citeyearpar{black1998} found evidence that
innovations designed to improve frequent feedback can bring substantial
learning gains. In addition, a number of studies observe that feedback,
when done well, can support learning
\citep{hattie2007power, carless2006}, especially for students with
little prior knowledge \citep{krause2009effects}. In concordance with
these observations, several scholars have argued for the need to raise
awareness among teachers on the usefulness of formative feedback
\citep{nicol2006formative, perera2008formative, irons2021enhancing}.

Impactful feedback should clarify what good performance is and encourage
positive motivational beliefs \citep{nicol2006formative}, it should be
provided in a timely manner, and be constructive and specific to the
student's work \citep{juwah2004enhancing, race2001using}. Providing
accurate feedback and grades to students' work can therefore be very
time consuming and become a struggle for (even moderately) large
classes. Data science courses are particularly affected, as they face
higher course enrollment numbers as one of their major challenges
\citep{NASEM}.

Automated grading and feedback tools have been proposed by many as a
possible solution \citep{galassi2021automated}. For types of assessment
where all ways of stating the correct solutions can be enumerated, like
multiple-choice, select-all and short-answer questions, automated
grading tools are available for assignments distributed using popular
online grading systems such as Gradescope \citep{singh2017} or learning
management systems (LMS) such as Canvas. Similarly, for computing
assignments where the correct solutions can be defined through a series
of if-else statements, several automated grading tools are currently
available, including the Python library \texttt{nbgrader}
\citep{nbgrader} for grading Jupyter notebooks, the R package
\texttt{learnr} \citep{learnr} for creating self-paced interactive
tutorials in teaching R and R packages, Otter-Grader \citep{otter} for
grading Python and R assignments, and the language-agnostic autograder
in Gradescope. These tools allow to give fast or even real-time
responses \citep[as recommended by][]{GAISE2016} to those assignments
for which, upon setup, providing grades and feedback can be done without
human judgement. However, they are not amenable for many recommended
assessment types, for which providing feedback remains challenging.

Indeed, open-ended assessments including lab reports \citep{GAISE2016},
semester long projects \citep{GAISE2016, finalprojects}, and writing
assignments \citep{woodard2020writing, johnson2016incorporating} cannot
be autograded, as enumeration of all possible correct or incorrect
approaches is not possible. These types of open-ended assignments are
known to provide opportunities for developing communication, critical
thinking, and creativity skills in addition to supporting statistical
knowledge \citep{garfield1999assessment}. Providing good feedback to
such assessments involves pedagogical choices and requires human
judgement, which brings two major challenges: a higher time-cost for
grading and more difficulty to grade consistently across students. These
challenges can affect the number of open-ended assignments and the
quality of feedback that can be provided, even for small and
medium-sized classes. Using a detailed rubric for grading can greatly
help with consistency
\citep{ragupathi2020beyond, timmerman2011development}, but it requires
time and can sometimes be overlooked during the grading workflow.

In this article, we consider grading workflows that can assist with
providing grades and good-quality feedback to data science and
statistics assignments that require human judgement to be assessed.
Gradescope \citep{singh2017} and the LMS Canvas, among other online
tools, both provide valuable options for grading open-ended assignments.
Particularly Gradescope, as reported by teachers from many STEM
disciplines \citep{garcia2015, reck2019, yen2020}, has made grading
easier and faster, with features such as rubric-based grading for
transparency and consistency, dynamic rubric creation and group
assignments, to name a few. However, data science and statistics
undergraduate classes often include assignments that involve computing
and may be completed, e.g., as R scripts \citetext{\citealp[see for
example][]{hu2022}; \citealp[and][]{dogucu2021}}, combination of
computing, data visualization and writing using, e.g., R Markdown and
Quarto files \citep[for instance][]{loy2019}, multiple of these files
(e.g., for a final project) and may be administered using GitHub for a
top-down approach to teaching Git \citep{beckman2021, fiksel2019using}.
These cases are not easily handled with Gradescope.

The contributions of this article are threefold. In
Section~\ref{sec-concept} we present the steps of a typical grading
workflow for data science open-ended assignments, highlighting the steps
that can be automated, and discussing pedagogical tools such as rubrics
and feedback, to build the concept of an automated grading workflow.
Next, in Section~\ref{sec-user}, we introduce the package
\textbf{gradetools} \citep{R-gradetools}, that implements an
automated-grading workflow within the RStudio Graphical User Interface
(GUI). This package enables efficient and consistent grading directly
within RStudio, as well as scalable yet individualized feedback
provision, and integrates with the existing R package \textbf{ghclass}
\citep{R-ghclass} to streamline feedback distribution for assignments
managed with GitHub. Following this, in Section~\ref{sec-developer} we
examine the key underpinnings of the gradetools package that allow it to
be an automated grading workflow. In doing so, we intend to provide data
science educators with ideas for improving their grading workflows, and
possibly developing new automated-grading workflow tools adjusted to
their own grading needs. Lastly, in Section~\ref{sec-discussion} we
summarize key points and discuss the implications of this work for the
statistics and data science education community.

\hypertarget{sec-concept}{%
\section{What is an automated grading workflow?}\label{sec-concept}}

The assessment at scale of assignments that cannot be auto-graded
requires automated grading workflows, that is, systems that automate all
or most repetitive grading tasks so as to reduce the time and effort
required for a grader to assess students' work and possibly provide
high-quality feedback.

A natural starting point for designing an automated grading workflow is
to outline the tasks that are executed in a typical grading workflow. We
can break down a grading process into three phases - (1) preparation,
(2) grading and providing feedback, (3) finalization - each with their
respective tasks. Table~\ref{tbl-workflow} lists these phases and groups
tasks into two types: \emph{pedagogical} and \emph{administrative}. The
former have a direct impact on what students learn from their grade and
their feedback, while the latter are related to the logistics involved
with the grading process.

\hypertarget{tbl-workflow}{}
\begin{table}
\caption{\label{tbl-workflow}Phases of a typical grading workflow and corresponding tasks. Tasks
colored in \textcolor{mygreen}{green} are
\textcolor{mygreen}{pedagogical}, in black are administrative. The tasks
listed in the grading phase need to be repeated for each submission to
grade. }\tabularnewline

\centering\begingroup\fontsize{11}{13}\selectfont

\begin{tabular}{>{\raggedright\arraybackslash}p{1.8in}>{\raggedright\arraybackslash}p{2.3in}>{\raggedright\arraybackslash}p{1.8in}}
\toprule
\textbf{1. Preparation \ \faFileImport} & \textbf{2. Grading and Feedback \ \faPenSquare} & \textbf{3. Finalization \ \faFileExport}\\
\midrule
Collecting students' assignments & Retrieving and opening a submissions & Uploading grade sheets on class's learning management system\\
\addlinespace
\textcolor{mygreen}{Setting up a rubric} & \textcolor{mygreen}{Assigning grade and writing feedback based on current rubric} & Returning grades and feedback to students\\
\addlinespace
Setting up a grade sheet & \textcolor{mygreen}{Updating rubric as needed} & \\
\addlinespace
 & Updating record of student corresponding to this submission on the grade sheet & \\
\addlinespace
 & Closing the submission & \\
\bottomrule
\end{tabular}
\endgroup{}
\end{table}

Automated grading workflows should automate administrative grading
tasks. These tasks tend to be repetitive and mostly do not require human
judgement during their execution; in fact, their automation can minimize
the occurrence of errors such as miscomputing the overall grade or
assigning a grade to the wrong student in the grade book.

Some pedagogical tasks - drafting and updating a rubric - always require
human judgement; other pedagogical tasks - evaluating a submission,
considering how the rubric should be applied to a given submission -
require human judgement for open-ended assignments. Even when they do
require human judgement, there are sources of repetitivity involved in
executing pedagogical tasks that an automated grading workflow can
automate.

Unless a class has very few students, most of the time required for
grading open-ended data science assignments is typically spent in Phase
2, that is, evaluating submissions and assigning grades and feedback.
Providing individualized feedback is especially time-demanding and may
often be sacrificed, even though it is extremely valuable for students
as discussed in Section~\ref{sec-intro}. Automated grading workflows can
scale provision of individualized feedback by leveraging the fact that
some feedback that is individualized to features of a student's
submission is in fact applicable to all students whose submissions
present the same features. To better illustrate this, the next
subsection outlines different types of feedback and explains how
automation may be achieved.

\hypertarget{feedback}{%
\subsection{Feedback types and automation}\label{feedback}}

Feedback can differ based on its applicability across students and
across questions. Table~\ref{tbl-feedback} distinguishes and exemplifies
six types of feedback, based on whether they are applicable to a single
or to multiple students and to a single question, to multiple questions
or to the entire assignment.

\hypertarget{tbl-feedback}{}
\begin{table}
\caption{\label{tbl-feedback}Examples of feedback that can be given to only a single or to multiple
students, and for only one question or for multiple questions (or
components). }\tabularnewline

\centering
\fontsize{11}{13}\selectfont
\begin{tabular}{>{\raggedright\arraybackslash}p{0.9in}>{\raggedright\arraybackslash}p{0.9in}>{\raggedright\arraybackslash}p{4.1in}}
\toprule
Student applicability & Question applicability & Example\\
\midrule
\cellcolor{gray!6}{multiple} & \cellcolor{gray!6}{single} & \cellcolor{gray!6}{When interpreting the slope coefficient make sure to use units of measurement.}\\
multiple & multiple & Please adhere to the Tidyverse style guide.\\
\cellcolor{gray!6}{multiple} & \cellcolor{gray!6}{entire assignment} & \cellcolor{gray!6}{Great job on this assignment!}\\
single & single & Recall our conversation about the p-value during office hour...\\
\cellcolor{gray!6}{single} & \cellcolor{gray!6}{multiple} & \cellcolor{gray!6}{The soft g letter (ğ) encoding is not displayed correctly on your output. In LaTeX try: \textbackslash u\{g\}.}\\
\addlinespace
single & entire assignment & Thank you for your note, Menglin. I am glad you had fun doing the assignment.\\
\bottomrule
\end{tabular}
\end{table}

Note that all of these feedback are \emph{individualized}, in the sense
that they are specific to a student's submission - rather than merely
stating what the correct solution is or summarizing how the whole class
did on the assignment. However, some of these feedback are
\emph{repeatable} - those that can be applied to multiple students -
while some of them are \emph{unique} because they are specific to a
feature that is present in a single submission.

Whether they are repeatable or unique to a submission, feedback can be
applicable to a \emph{single question} of the assignment (or component),
applicable to \emph{multiple questions} (or components), or be
\emph{general} feedback that refer to how a student did overall on the
assignment. In our experience, we found that multiple-question,
single-student feedback is rarely needed but we commonly encounter the
other feedback scenarios.

An automated grading workflow should expedite the provision of
repeatable feedback across different questions and students. It should
also facilitate providing unique feedback on the fly. One way of scaling
the evaluation of submissions and the assignment of grades and feedback
is by setting up a rubric that has an item for each encountered feature
of students' assignments and that, for each item, indicates both its
associated feedback and its associated score (that is, a number of
points to remove, or add, to a student's grade when their work presents
this feature). This will be further discussed in
Section~\ref{sec-developer}.

Given the rubric and a selection of rubric items prepared by the grader,
a system can then be designed to update the grade sheet and write
individualized feedback to the assignment that is being graded. An
automated grading workflow should also facilitate updating the rubric
dynamically, as previously-unobserved assignments' features are
encountered that may occur in other submissions yet to be assessed.

The R package gradetools is an automated-grading-workflow system
designed around these ideas, for executing Phase 2 tasks within RStudio.
The next section shows what this system looks like from the user
perspective.

\hypertarget{sec-user}{%
\section{Introducing gradetools}\label{sec-user}}

\hypertarget{motivating-grading-scenario}{%
\subsection{Motivating grading
scenario}\label{motivating-grading-scenario}}

Earlier functions of the gradetools package were first developed to
assist with grading assignments of the Stats 68 class - Statistical
Computing and Exploratory Data Analysis with roughly 60 students
enrolled. Since its creation as a package, we also used gradetools
extensively in our Stats 6 - Introduction to Data Science course with
around 120 students. The courses share a similar computational
structure. Every week students receive and submit assignments through
individual GitHub repositories. Formerly, the assignments generally
consisted of a R Markdown file (.Rmd), with questions involving a
combination of coding and text (e.g., running a model and interpret its
results), or occasionally R scripts (.R). Currently, we utilize Quarto
files (.qmd) in our courses.

Teaching students good coding practices and reproducible workflows via R
Markdown/Quarto and GitHub have been key learning goals of
computationally rigorous courses that we teach. In these courses, we
assesses the raw .Rmd/.qmd files rather than their rendered pdf or HTML
files. In addition, students work on projects that have multiple files
with directories consisting of multiple folders. For instance, a project
folder has subfolders consisting of data, project proposal, and
presentation with each subfolder having multiple files.

For such teaching settings, we had a few options to consider. The first
one was the LMS Canvas. Grading on Canvas can be efficient when each
student submits a single pdf file for an open-ended assignment. The
interface allows for annotation and use of rubric. Canvas did not meet
our grading needs since students submitted multiple files (e.g., a
dataset they found and an Rmd file) and Rmd files could not be
displayed. We did not consider downloading students' submissions as a
zipped folder and opening one-by-one as that would defeat the goal of
efficient grading.

Another option was autograders including the aforementioned packages or
Gradescope's automated grading feature. These options did not allow for
assessing higher-levels of thinking that we wanted to assess. Lastly,
Gradescope also has an online submission feature, which in its capacity
is similar to an LMS. Contrary to Canvas, Gradescope does display
markdown documents. However, this also limits student's submission to a
single file submission at a time and cannot help instructors assess a
full project in a typical folder structure.

We believe that in similar computationally rigorous statistics and data
science courses, students' files are best suited to be evaluated in a
GUI such as RStudio, where a grader can look at the raw file and
simultaneously, if needed, its rendered version. To avoid the need of
switching between different applications used to maintain a grade sheet
and manage a rubric, we created the gradetools package to carry out
grading within RStudio. With this package we also automated repetitive
parts of the grading workflow, and we automated writing a rich,
individualized feedback file for each student that could then be shared
with them.

\hypertarget{sec-grading-ex}{%
\subsection{Grading example with the gradetools package in
RStudio}\label{sec-grading-ex}}

To better understand the process of grading with gradetools, we will
walk through a simple example of grading a quiz with two questions. In
this example we begin with a properly formatted class roster, quiz
rubric, and student submissions. Details on gradetools's format
requirements can be found in Section~\ref{sec-developer} and in our
introductory vignette
(\url{https://federicazoe.github.io/gradetools/articles/a-grading-with-gradetools.html}).

\textbf{Grading in action}

\begin{figure}

{\centering 

\includegraphics[width=145mm,height=\textheight]{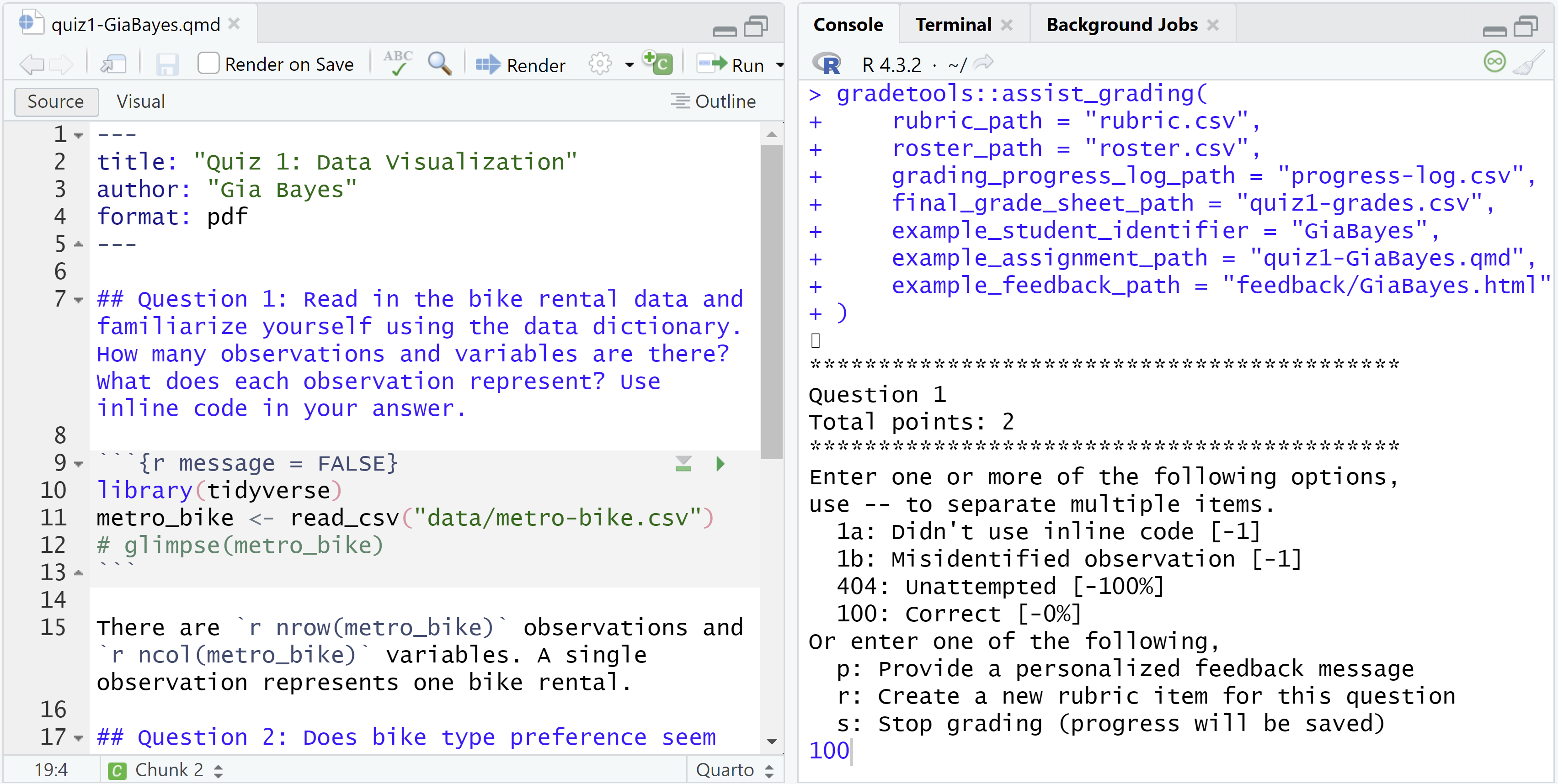}

\includegraphics[width=145mm,height=\textheight]{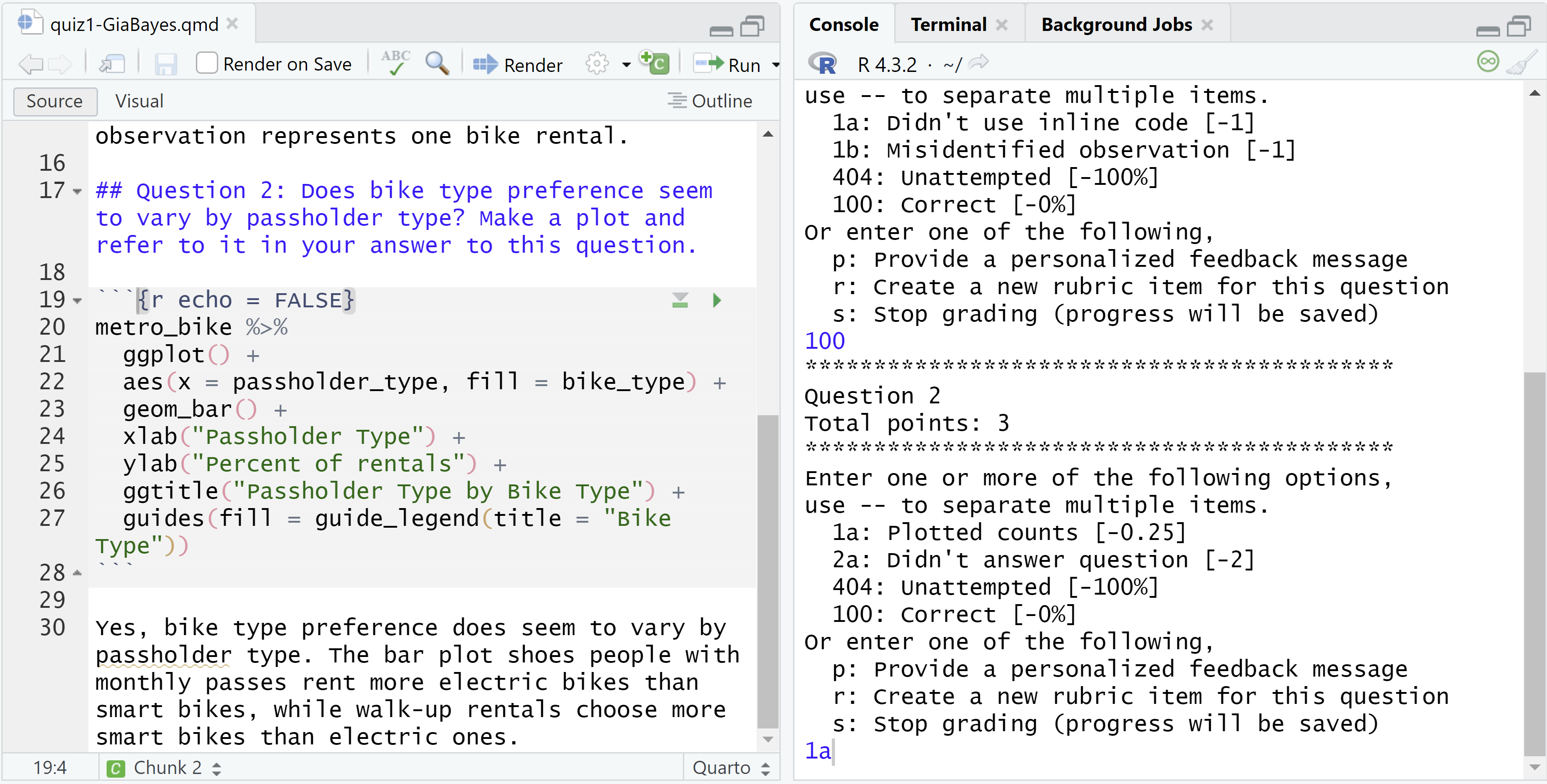}

\includegraphics[width=145mm,height=\textheight]{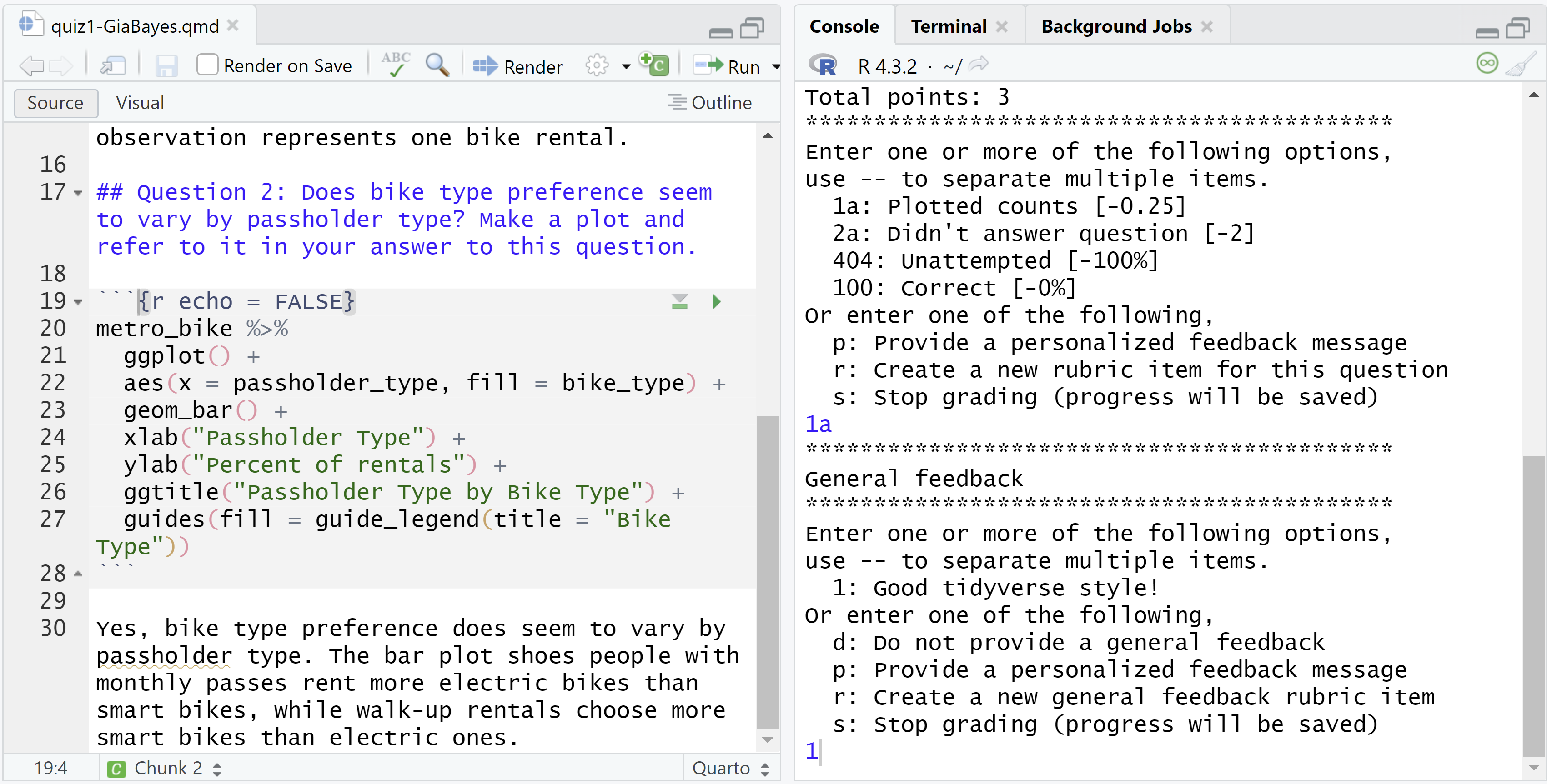}

}

\caption{\label{fig-grading-example-student1}Example of grading a quiz
using gradetools in RStudio. Each image shows the grading of a component
of a student's quiz. This quiz belongs to the first student on the
roster, Gia Bayes.}

\end{figure}

To begin the process, the gradetools package is loaded and the grading
function, \texttt{assist\_grading()} is called - this is shown in the
first image of Figure~\ref{fig-grading-example-student1}. This function
requires the locations of the roster, rubric, and submissions, as well
as where to write the grading progress log, grade sheet, and feedback
files. The function call triggers the grading of the first student on
the roster and opens their submission automatically. Based on the
provided location of the submission and on the desired location for the
feedback file of a \emph{single} student, gradetools determines the file
path for submission and feedback files of all students in the roster.
Grading is an interactive process where the grader is prompted in the
console to grade a submission according to the provided rubric.

Figure~\ref{fig-grading-example-student1} displays three screenshots,
each showing the student's quiz on the left, and on the right in the
console is the rubric prompting and selection for a component of the
quiz. The first screenshot shows the grading function call (on the
right), which begins the grading process by automatically opening the
first student's assignment submission (on the left) and prompting the
grader to grade the first question on the quiz according to rubric
options (on the right). This student, Gia Bayes, has answered both parts
of Question 1 correctly so the ``Correct'' rubric item is selected,
removing zero points, by entering ``100'', the corresponding prompt
code, into the console. This example is using a negative grading scheme,
where each rubric item is associated with points to remove from the
total number of points the question is worth. After the code is entered
into the console, the user is then prompted to grade the next question
in the rubric. Note that there are additional options - providing a
personalized feedback message, creating a new rubric item, and
terminating the grading process - and these options will be discussed
later.

Moving onto question 2, displayed in the second screenshot in
Figure~\ref{fig-grading-example-student1}, the student was asked to
produce a plot and use it to compare within-group patterns. Overall the
student addressed this question, but plotted counts instead of
proportions. We want to convey that plotting proportions within groups
would visualize all group patterns, while plotting counts hinders
comparison of patterns within the groups with relatively small counts.
The corresponding rubic item is the first available, ``Plotted counts''
which deducts 0.25 points, and is selected by supplying ``1a'' to the
console.

After all the questions have been graded, the user is prompted to
provide feedback for how the student did on the quiz overall, called
general feedback. This rubric has one general feedback rubric option,
which commends the student's adherence to the tidyverse style guide. Gia
Bayes did a great job using tidyverse style so this option is selected
by entering ``100'' into the console.

\begin{figure}

{\centering 

\includegraphics[width=145mm,height=\textheight]{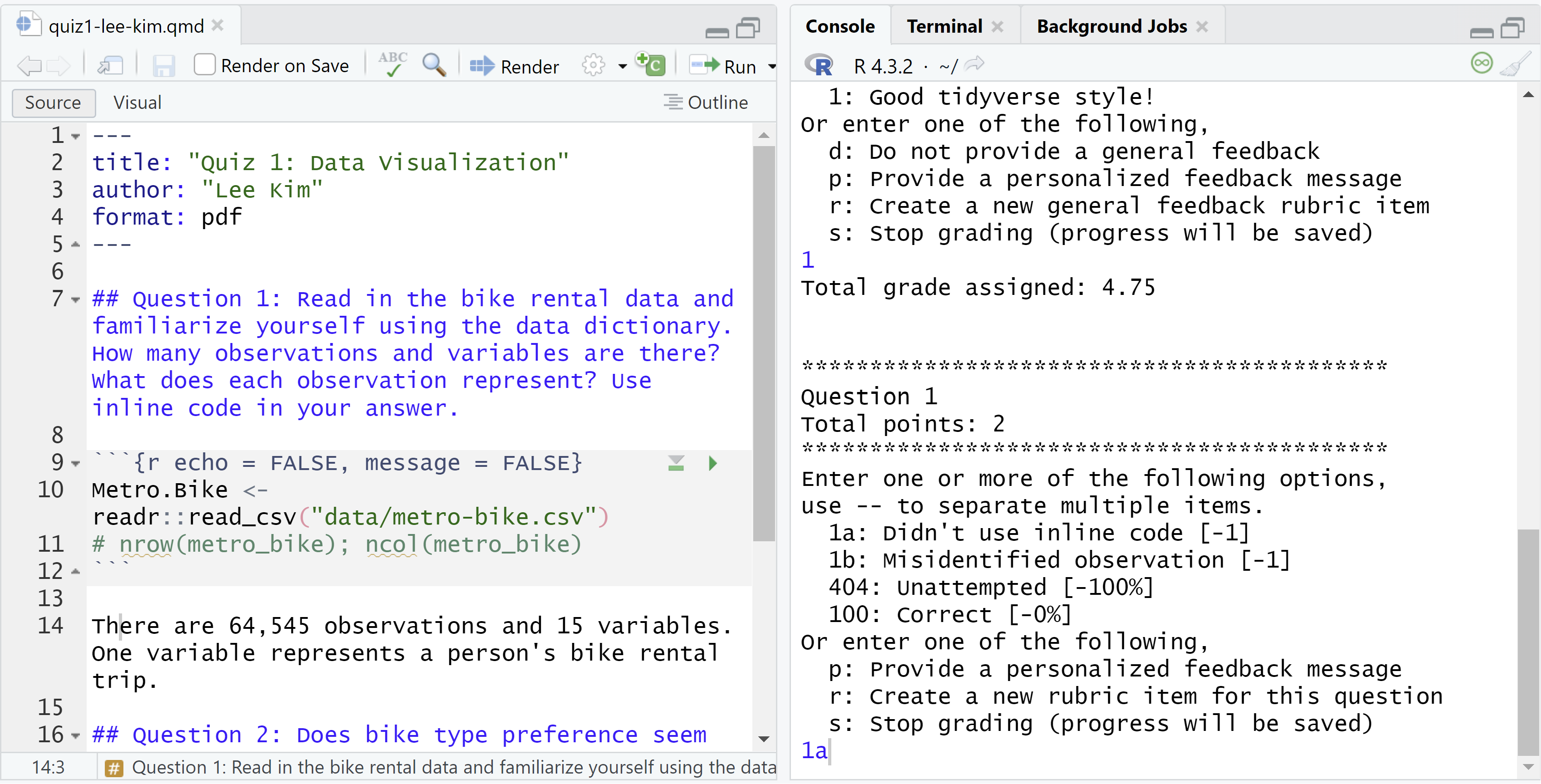}

\includegraphics[width=145mm,height=\textheight]{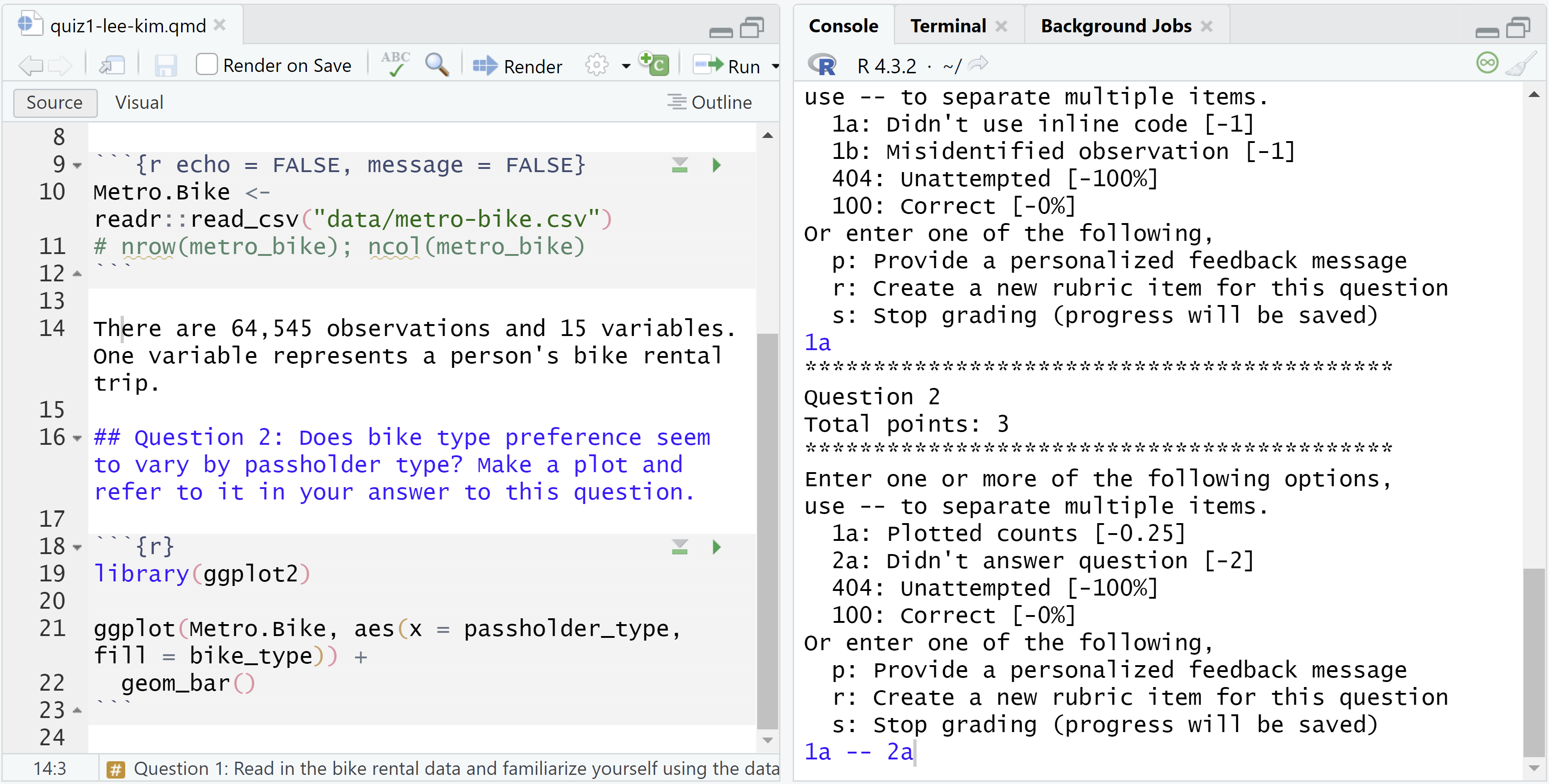}

\includegraphics[width=145mm,height=\textheight]{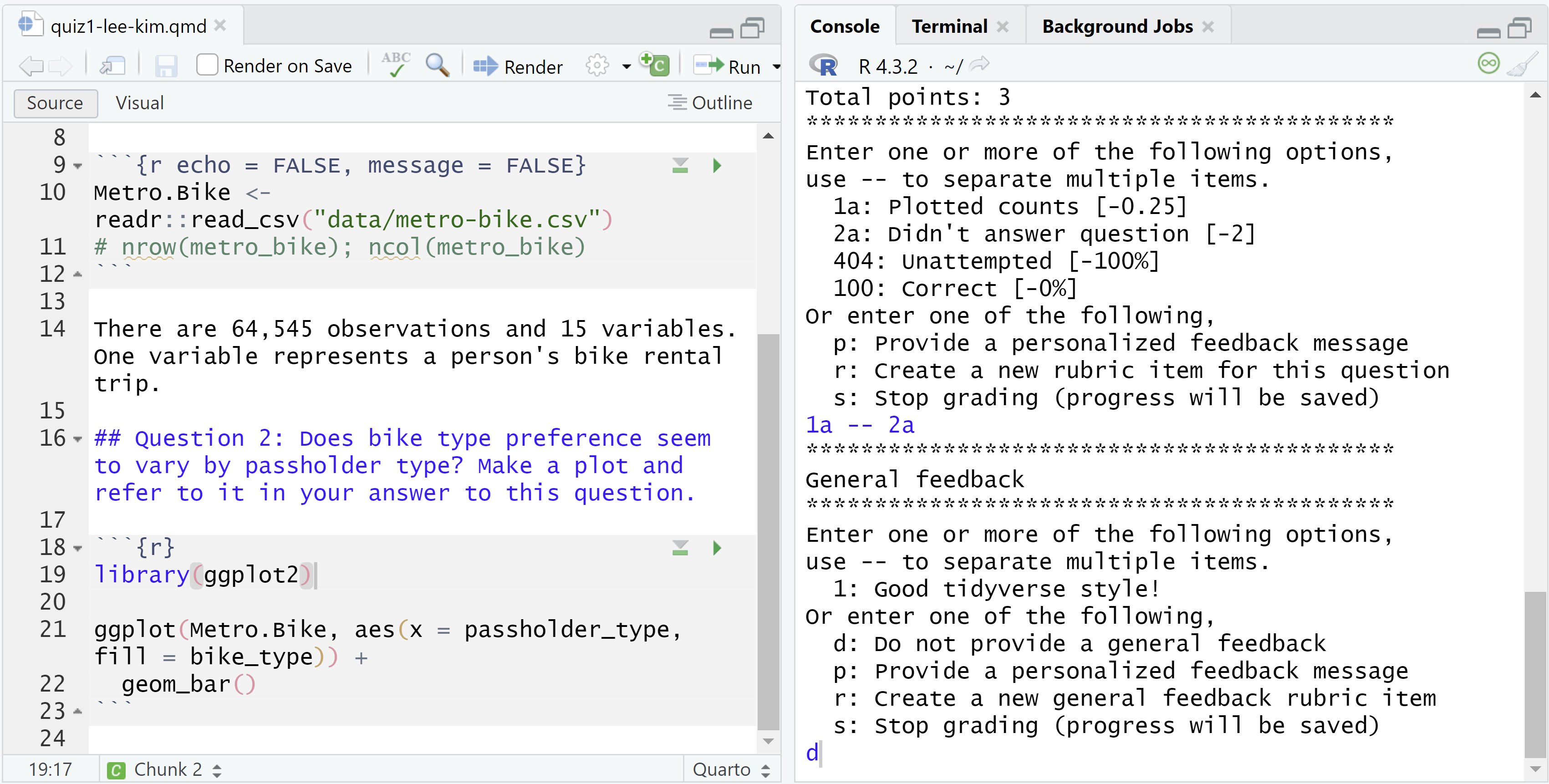}

}

\caption{\label{fig-grading-example-student2}Continuation of example of
grading a quiz using gradetools in RStudio. Each image shows the grading
of a component of a student's quiz. This quiz belongs to the second
student on the roster, Lee Kim.}

\end{figure}

Once all questions have been graded, the submission automatically
closes, the student's overall quiz grade is displayed, and grading
continues with the next student in the roster, Lee Kim. As shown in
Figure~\ref{fig-grading-example-student2}, Lee's quiz is automatically
opened and the grader is prompted to grade question 1. Lee correctly
identified the number of observations and variables, but did not use
inline code in their answer, so the rubric item ``Didn't use inline
code'' is applied by entering the prompt code ``1a'', which will deduct
1 point from the question's total of 2 points. For question 2 we see
that the student did a bar plot of counts, like the first student we
graded, and they did not provide an explicit answer to the question
``Does bike type preference seem to vary by passholder type?''. These
mistakes each correspond to a different rubric item. Multiple rubric
items are applied by using ``--'' to separate prompt codes. After
entering ``1a -- 2a'', the grader is asked to provide general feedback
for Lee's quiz. Tidyverse style was not used for naming the bike rental
data, so we may choose to avoid providing a general feedback by entering
``d'' into the console.

\textbf{Grading outputs}

\begin{figure}

\begin{minipage}[t]{0.50\linewidth}

{\centering 

\raisebox{-\height}{

\includegraphics[width=80mm,height=\textheight]{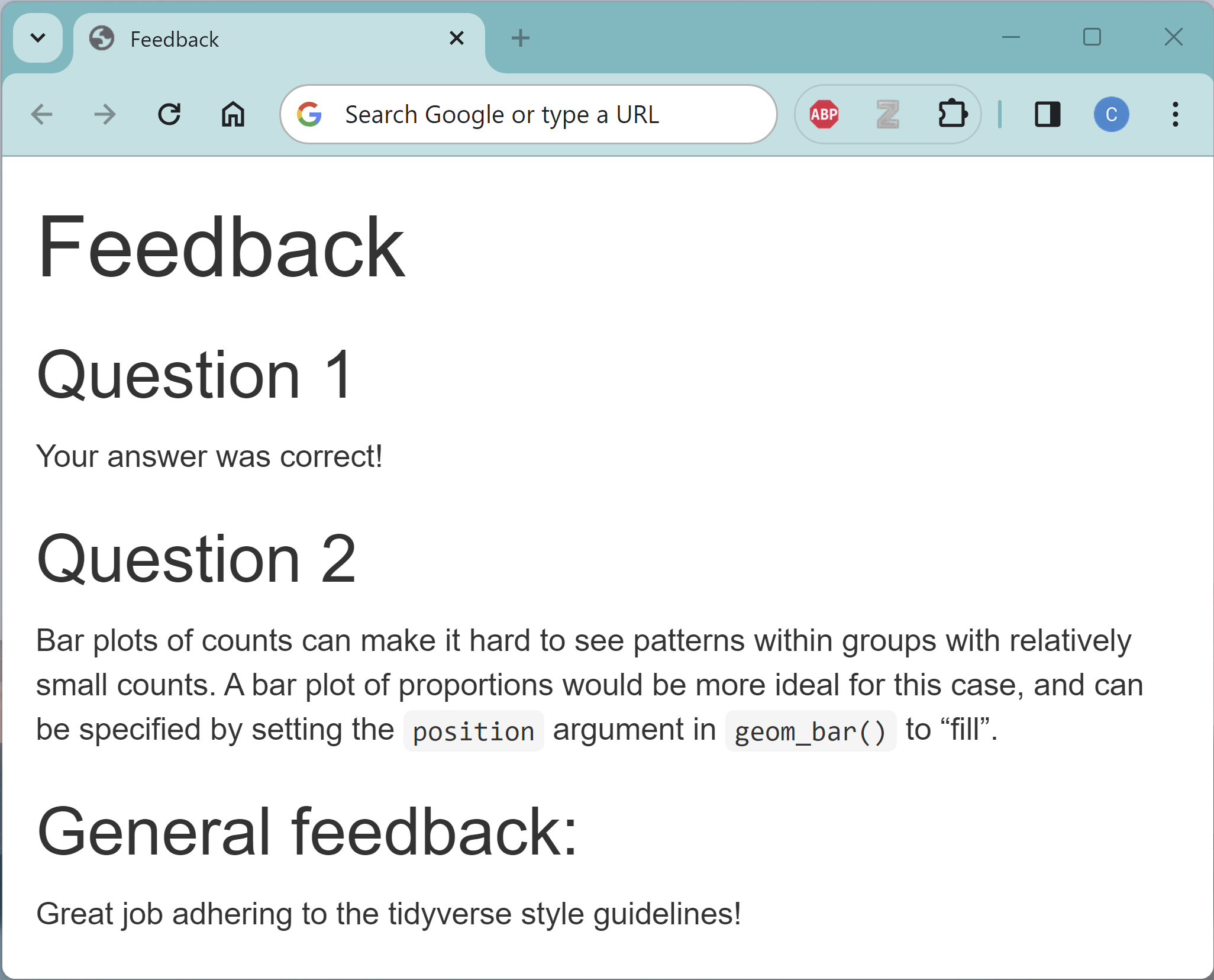}

}

}

\subcaption{\label{fig-grading-example-feedback1}Feedback for first
student, Gia Bayes}
\end{minipage}%
\begin{minipage}[t]{0.50\linewidth}

{\centering 

\raisebox{-\height}{

\includegraphics[width=80mm,height=\textheight]{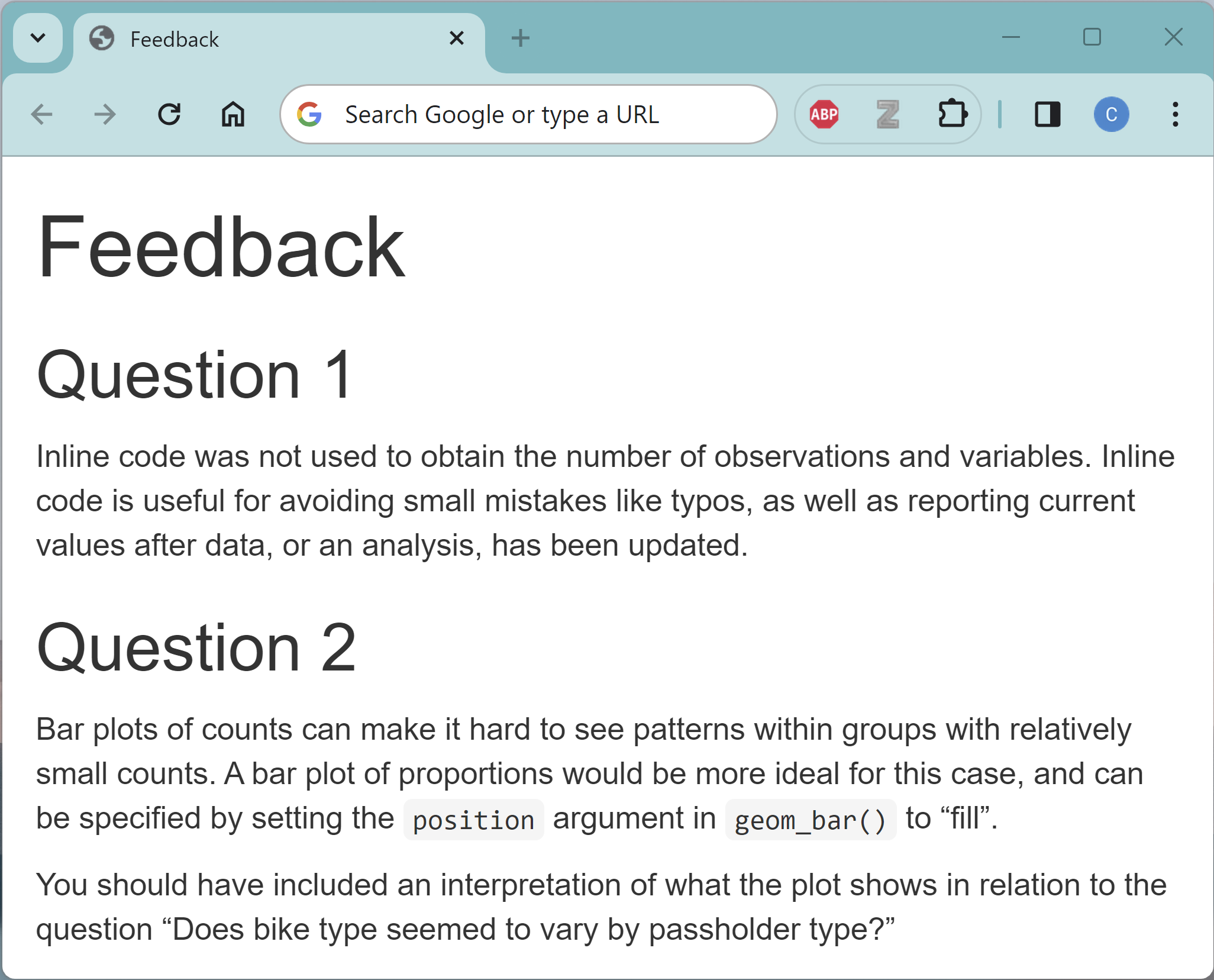}

}

}

\subcaption{\label{fig-grading-example-feedback2}Feedback for second
student, Lee Kim}
\end{minipage}%
\newline
\begin{minipage}[t]{0.50\linewidth}

{\centering 

\raisebox{-\height}{

\includegraphics[width=80mm,height=\textheight]{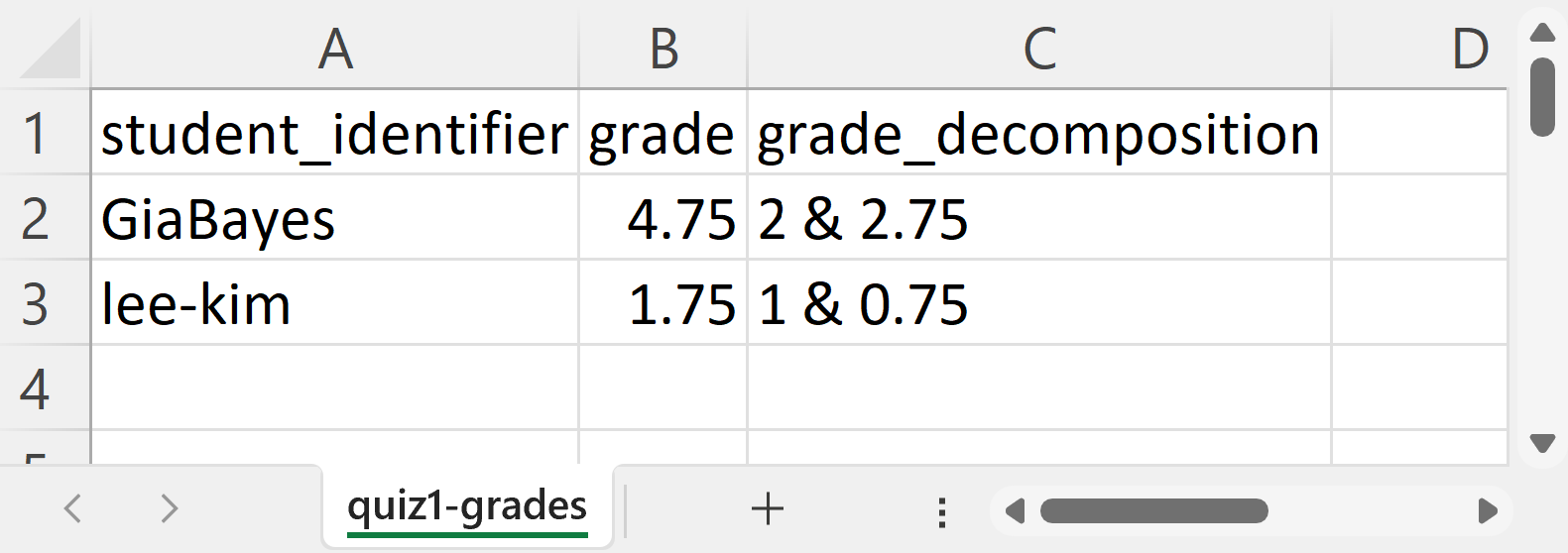}

}

}

\subcaption{\label{fig-grading-example-grade-sheet}Grade sheet}
\end{minipage}%

\caption{\label{fig-grading-example-outputs}Outputs of grading: feedback
files and grade sheet corresponding to example of grading a quiz using
gradetools in RStudio.}

\end{figure}

To conclude this example grading session we will proceed as if all
remaining quizzes have been graded. Upon completion of grading, or
termination of the grading process, a final grade sheet is created and
feedback files are automatically written from the feedback associated
with the applied rubric items for each student, displayed in
Figure~\ref{fig-grading-example-outputs}.

Figure~\ref{fig-grading-example-grade-sheet} shows the grade sheet for
this quiz, with student identifiers obtained from the class roster,
total grade, and points earned for each question, separated by an
ampersand. The grade sheet could now be formatted and uploaded to a LMS
and the feedback files could be distributed to students. While grade
sheets are common in practice, feedback files may be less familiar.

The first student's feedback file displayed in
Figure~\ref{fig-grading-example-feedback1} contains feedback associated
with the rubric items selected when grading this student. Notice that
the prompt message for the rubric item applied for question two,
``Plotted counts'', is different than the feedback message written to
the file. Both messages were specified for that rubric item in the
rubric file, one meant as a concise summary for the grader and the other
as a more thorough note for the student. The association of feedback
messages with rubric items is a key aspect of gradetools which avoids
redundancy of retyping feedback for the same mistake, while creating
feedback documents that are specific to each students' performance on
the assignment. For example, both students made the same mistake of
making a bar plot of counts for question two, so they both have the same
corresponding message written in their feedback file. The second student
made the additional mistake of not explicitly answering the question
posed in question 2, so they have an additional message in their
feedback.

\textbf{Dynamic rubric editing}

In this grading example the rubric already had rubric items for all
instances we encountered. When that is not the case, the grader would
likely want to add rubric items as they grade and encounter new
responses. This is possible with gradetools by entering ``r'' into the
console instead of specifying an available rubric item. Another
possibility is that the grader may want to edit a preexisting rubric
item, e.g., change the number of points possible for a question or the
feedback for a rubric item. Whenever the grader re-runs the grading
function with an updated rubric, all feedback files and the grade sheet
are updated to reflect the latest rubric version.

\textbf{Unique on-the-fly feedback}

Another option that was not showcased in the previous graing example is
the ability to write feedback message unique to a student - that is, a
``single-student'' type of feedback in the terminology of
Table~\ref{tbl-feedback}. By entering ``p'' into the console while
grading, the user can enter a note to be written to the feedback file
not associated with any points. For example for the second student,
question 2, we could have left a note in their feedback telling the
student they could receive partial credit if they resubmit their
assignment with an interpretation of the plot in response to the
question.

\textbf{Fixing grading mistakes}

Lastly, in this example we did not make any mistakes, but mistakes can
happen in reality. Grading can be stopped at any time by entering ``s''
into the console. Doing so will end the grading process and all grading
progress will be maintained through the grading progress log produced.
The function \texttt{assist\_regrading()} can be used to regrade
specified students and questions (see the vignette for step-by-step
instructions at
\url{https://federicazoe.github.io/gradetools/articles/b-regrading-with-gradetools.html}).

\hypertarget{grading-scenarios}{%
\subsection{Grading scenarios}\label{grading-scenarios}}

\begin{figure}

{\centering \includegraphics[width=1\textwidth,height=\textheight]{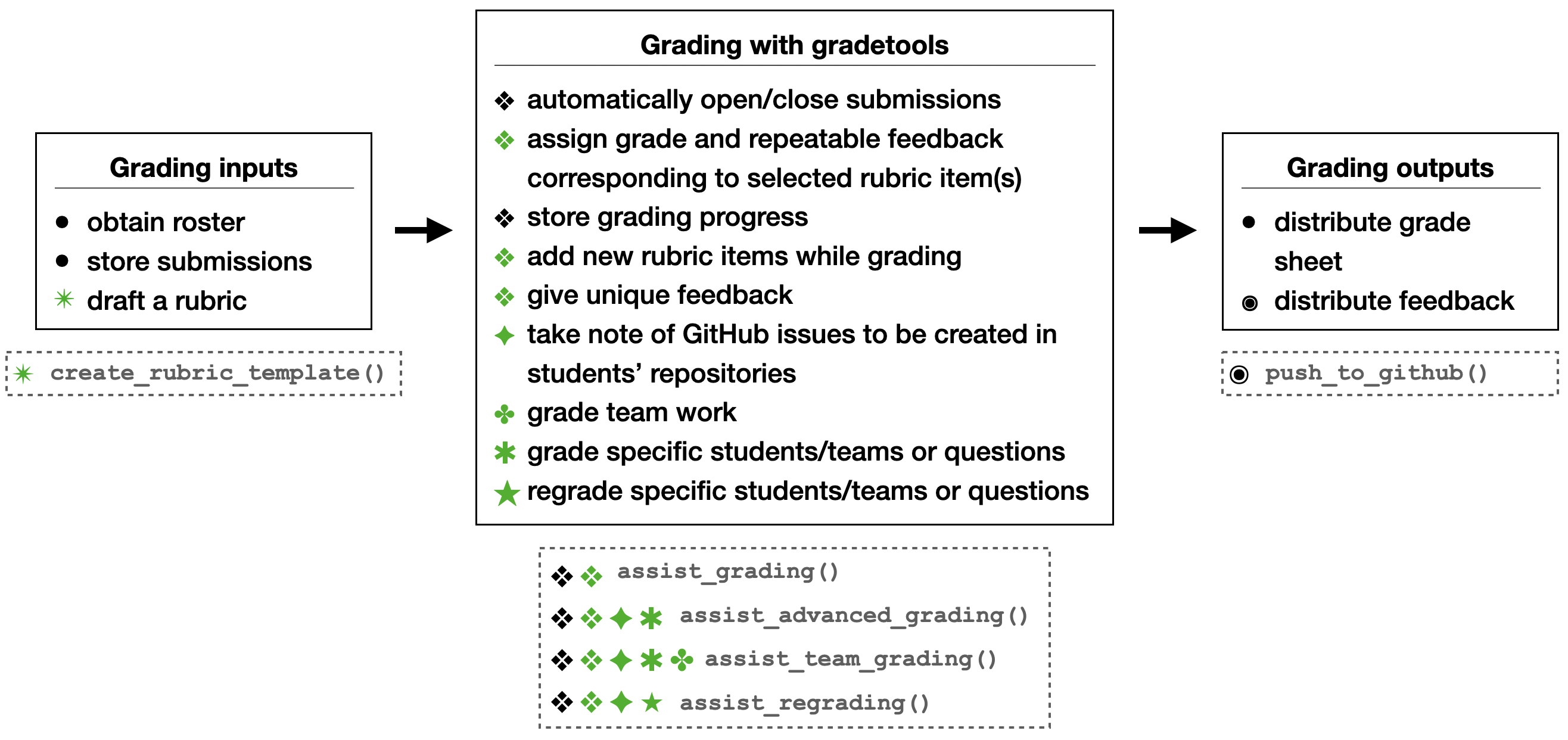}

}

\caption{\label{fig-gradetools-diagram}Diagram of automated grading
workflow using gradetools in the grading step. Administrative tasks in
black and \textcolor{mygreen}{pedagogical} tasks in
\textcolor{mygreen}{green}. Special bullet shapes map tasks that
gradetools assists with to the name of the functions that support them.
\emph{Repeatable} and \emph{unique} feedback refer to the concepts
defined in Section \ref{feedback}.}

\end{figure}

The grading function \texttt{assist\_grading()} has the core grading
functionalities (later summarized in
Figure~\ref{fig-gradetools-diagram}) and is useful for users with
limited R knowledge. The previous grading example demonstrated using
gradetools to grade a single file per student. We will now discuss other
grading scenarios and their respective gradetools functions.

\textbf{Grading projects}

Sometimes assignments involve multiple files to grade, for example a
final project that includes a README file to describe where the data was
obtained and a Quarto file that generates a presentation. Assignments
that involve multiple submissions can be handled by gradetools, by
providing a vector of file paths for a student's submission, instead of
just a single location. Another useful ability for grading projects in
RStudio with gradetools is the option to render documents while grading,
so the raw and rendered files can be viewed at the same time. Examples
of these features for grading projects can be found in our comprehensive
vignette
(\url{https://federicazoe.github.io/gradetools/articles/e-comprehensive-example.html}).

\textbf{Grading team assignments}

We use \emph{team grading} to refer to the case when multiple students
share a submission and grade. The \texttt{assist\_team\_grading()}
function allows for grading team assignments and functions similarly to
\texttt{assist\_grading()}. The only additional requirement is a column
in the roster denoting what team each student is on. The grading process
for team assignments results in a single feedback file and grade for
each team. A vignette for team grading can be found at
\url{https://federicazoe.github.io/gradetools/articles/c-extended-capability-teams.html}.

\textbf{Grading assignments managed through GitHub}

Assignments can be managed through GitHub, teaching students
reproducibility practices and version control, but collecting
assignments from GitHub, grading, and returning feedback to student can
be time consuming without the appropriate tools, such as the R package
ghclass. Our package complements the streamlined collection of GitHub
repositories with ghclass, by allowing the noting of issues while
grading with \texttt{assist\_advanced\_grading()} or
\texttt{assist\_team\_grading()}, and allowing the user to push feedback
and create issues on GitHub using \texttt{push\_to\_github()}. A
vignette on using gradetools with assignments managed through GitHub can
be found at
\url{https://federicazoe.github.io/gradetools/articles/d-extended-capability-github.html}.

\textbf{Multiple graders}

When an instructor has helpers, such as teaching assistants, the grading
load can be split accross multiple people. This may mean the students
submissions are partitioned between the graders (e.g.~grader 1 the first
20 students and grader 2 the last 20 students), or the questions for all
submission may be partitioned between the graders (e.g.~grader 1 grades
the odd questions and grader 2 the even). The functions
\texttt{assist\_advanced\_grading()} and
\texttt{assist\_team\_grading()} allow for the user to specify which
students and questions are to be graded, with the default of all
students and questions. Simultaneous grading with gradetools would
result in different grading log files for each grader (see
Section~\ref{sec-maintain-grade} for details on this file). Merging
grading log files would then need to be implemented by the grading team,
as gradetools does not currently include a function for it.

\hypertarget{considerations-for-adoption}{%
\subsection{Considerations for
adoption}\label{considerations-for-adoption}}

When considering adopting gradetools as your automated-grading-workflow
assistant, it is important to take into account your grading scenarios
and if gradetools is compatible, and weigh the advantages against the
learning curve. This package was made with coding and report scripts for
data science assignments in mind for moderately sized classes, where
there are no grading packages or software that we are aware of. But
gradetools can also be helpful for grading scenarios beyond its original
purpose, especially for teachers at institutions that do not pay for
grading software such as Gradescope.

A key consideration for adopting gradetools, or any other software, is
the learning curve. Minimal R knowledge is required for gradetools,
since the user only needs to know how to call a function from a package,
but an understanding of file paths is necessary since the arguments for
the grading functions are almost all file paths. The biggest challenge
in adoption would be learning the details of how the rubric must be
formatted and the file naming conventions in order to use gradetools.
These requirements are detailed in our introductory vignette
(\url{https://federicazoe.github.io/gradetools/articles/a-grading-with-gradetools.html}),
and their purpose is discussed in Section~\ref{sec-developer}. Once a
user has successfully called the grading function, the grading process
is straightforward, as exemplified in Section~\ref{sec-grading-ex}.

The time to grade is an important decision when creating assignments and
deciding on a grading workflow. Grading open ended questions can be time
consuming, and gradetools speeds up the process by automating repetitive
tasks, but still requires considerable time relative to assignments that
could be autograded. For large classes and limited teaching staff, the
time gain for using gradetools may still not be enough to allow grading
many open-ended assignments throughout the quarter, and a combination of
frequent quizzes autograded with another tool (e.g., Online assignments
in Gradescope) with few open-ended assignments graded with gradetools
may be considered.

We restricted the scope of gradetools to mostly the grading stage,
leaving collection of student submissions, distribution of feedback
files, and uploading grade sheets mostly beyond the scope of our
package, with the exception of GitHub-compatible functionalities. Our
package could be combined with other available software, such as a
package to distribute feedback files through email or packages which can
aid in the collection of student submissions, in order to speed up other
components of your grading workflow.

\hypertarget{sec-developer}{%
\section{Underpinnings of automation in
gradetools}\label{sec-developer}}

In this section we provide more details on how gradetools is designed.
The package's vignettes that can be found at
\url{https://federicazoe.github.io/gradetools/} are the best
documentation to get started to \emph{use} gradetools functionalities.
Instead, this section is for the reader who would like to understand how
gradetools provides those functionalities, and perhaps may consider to
extend gradetools or develop new automated grading-workflow tools to
better suit their grading needs.

In the next subsections we discuss the main strategies adopted by
gradetools for automating repetitive tasks encountered in Phase 2 of
grading (i.e.~assigning grades and feedback), that were outlined in
Table~\ref{tbl-workflow}. Specifically, we consider automation with: (1)
retrieving, opening and closing submissions; (2) assigning grade and
feedback based on rubric; (3) maintaining the grade sheet. For each of
these tasks, we are going to see that the way in which they are
automated is tied to specific choices made during Phase 1 of grading
(i.e.~grading preparation).

\hypertarget{sec-open-submis}{%
\subsection{Retrieving, opening and closing
submissions}\label{sec-open-submis}}

With many submissions, considerable time may go into retrieving, opening
and closing submission file(s) without automation of these tasks. As
seen in the grading example of Section~\ref{sec-grading-ex}, these are
tasks that are automated by gradetools. To automatically identify the
student corresponding to each submission, some assumption must be made
on the way in which students assignments are collected (or better,
stored). The solution that we adopted with gradetools is to assume that
students have a unique student identifier, specified in the roster, and
that this identifier is present (at least once) in the student's
assignment file path and is the only part unique to that student's
assignment file path. For example, if the first quiz consists of a
single Quarto file, all submissions may be stored in a folder named
\texttt{quiz1} and the file names may be \texttt{quiz1-GiaBayes.qmd},
\texttt{quiz1-lee-kim.qmd}, etc. Thanks to this structure, at grading
time, the grader needs to provide the student identifier and assignment
file path for only one student in the roster. For example, a grader can
call \texttt{assist\_grading()} with inputs
\texttt{example\_student\_identifier\ =\ \textquotesingle{}GiaBayes\textquotesingle{}}
and
\texttt{example\_assignment\_path\ =\ \textquotesingle{}quiz1/quiz1-GiaBayes.qmd\textquotesingle{}},
and provide the location of the roster where the other student
identifiers are listed. Then, the template \texttt{quiz1/quiz1-} +
student identifier + \texttt{.qmd} is assumed for all assignment file
paths, and gradetools is able to iterate through opening and closing all
submissions.

An assignment made of multiple files can be handled in a similar way.
For example, if the first quiz consisted of two files (e.g., a Quarto
document with a data analysis and a R script with some functions) all
submissions may be stored in a folder named \texttt{quiz1} and the file
names for GiaBayes may be \texttt{analysis-GiaBayes.qmd} and
\texttt{functions-GiaBayes.R}, those for lee-kim may be
\texttt{analysis-lee-kim.qmd} and \texttt{functions-lee-kim.R}, etc. As
another example, both files could have the same name but be stored in a
folder with naming specific to the student, e.g., files
\texttt{analysis.qmd} and \texttt{functions.R} may be located in folders
\texttt{quiz1-GiaBayes}, \texttt{quiz1-lee-kim}, etc. In the above
cases, when grading we could provide
\texttt{example\_student\_identifier\ =\ \textquotesingle{}GiaBayes\textquotesingle{}}
and the vector
\texttt{example\_assignment\_path\ =\ c(\textquotesingle{}quiz1/analysis-GiaBayes.qmd\textquotesingle{},\ \textquotesingle{}quiz1/functions-GiaBayes.R\textquotesingle{})}
for the former case or
\texttt{example\_assignment\_path\ =\ c(\textquotesingle{}quiz1-GiaBayes/analysis.qmd\textquotesingle{},\ \textquotesingle{}quiz1-GiaBayes/functions.R\textquotesingle{})}
for the latter case.

Provided that student submissions are stored with these regular file
paths, besides iterating through all submissions, gradetools can also
retrieve the file(s) of one or more \emph{specific} students, for
example because we wish to grade or re-grade only their submissions. The
functions \texttt{assist\_advanced\_grading()} or
\texttt{assist\_regrading()} have optional arguments, respectively
\texttt{students\_to\_grade} and \texttt{students\_to\_regrade}, that
can be set to a single string or to a vector of strings specifying the
identifiers of the students we wish to (re-)grade.

\hypertarget{sec-assign-grade}{%
\subsection{Assigning grade and feedback based on
rubric}\label{sec-assign-grade}}

\begin{figure}

\begin{minipage}[t]{\linewidth}

{\centering 

\raisebox{-\height}{

\includegraphics[width=100mm,height=\textheight]{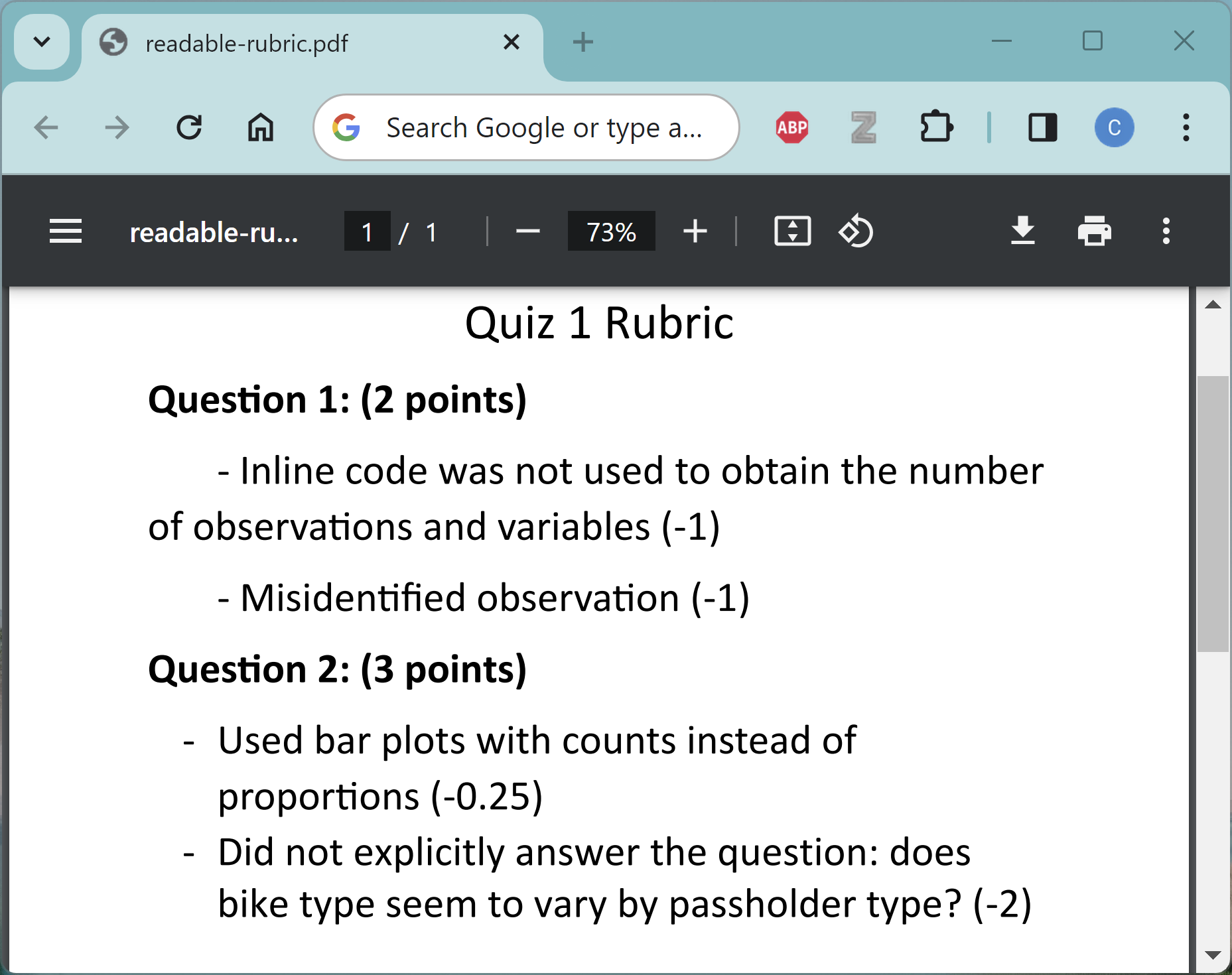}

}

}

\subcaption{\label{fig-grading-example-readable-rubric}Unformatted
rubric}
\end{minipage}%
\newline
\begin{minipage}[t]{\linewidth}

{\centering 

\raisebox{-\height}{

\includegraphics[width=160mm,height=\textheight]{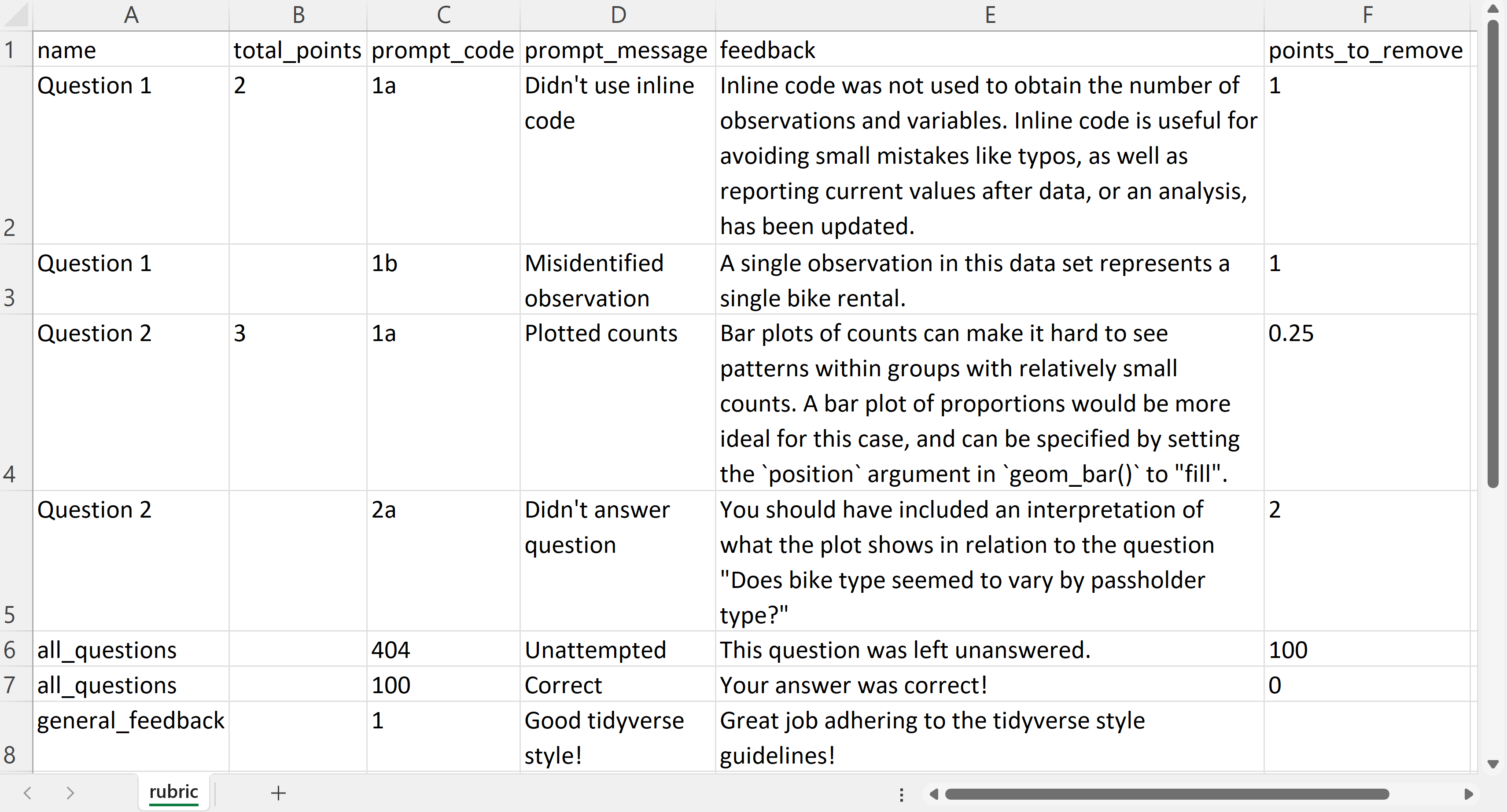}

}

}

\subcaption{\label{fig-grading-example-formatted-rubric}Formatted
rubric}
\end{minipage}%

\caption{\label{fig-grading-example-rubrics}Rubric from grading example
in its readable form and formatted to be compliant with gradetools
rubric requirements.}

\end{figure}

Once the submission file(s) of a students have been retrieved and
opened, we need to read (i.e., evaluate) the students' work. For each
component of the assignment, we must choose which items from the rubric
we want to apply, possibly add new items to the rubric, and ideally give
the student some qualitative feedback specific to their work. Without a
system like gradetools, these tasks may be extremely time and energy
consuming, as they require to simultaneously refer to and possibly edit
multiple grading files (the submission files, the rubric, the grade
sheet and some file where we write feedback). As illustrated with the
grading example of Section~\ref{sec-grading-ex} (e.g., see
Figure~\ref{fig-grading-example-student1}), the strategy we adopted with
gradetools enables to visualize the available rubric items directly in
RStudio, next to the open file(s), and only requires the grader to enter
one or more short prompts corresponding to the rubric items to be
applied to the submission. In this process, on the back end, the grade
sheet gets updated and a feedback file gets written for each student
that is graded.

The key to this automation is setting up a rubric that has all the
necessary information to (i) make grading prompts, such as those shown
in Figure~\ref{fig-grading-example-student1} and
Figure~\ref{fig-grading-example-student2}; and, based on the choices
made by the grader, (ii) compute the grade and (iii) assign personalized
feedback. Without gradetools, when grading the example assignment from
Section~\ref{sec-user} a grader may use a rubric such as the one
displayed in Figure~\ref{fig-grading-example-readable-rubric}. This
rubric breaks down the assignment into components (in this case,
questions) and, for each component, specifies its assigned points (e.g.,
2 points for Question 1). Each item of the rubric represents a potential
error (e.g., Mispecified observation) and notes the points to remove for
that error (e.g., -1). The rubric required by gradetools encodes all
this information, plus additional information to make rubric prompts and
write feedback. Figure~\ref{fig-grading-example-formatted-rubric}
displays the rubric provided to gradetools in the examples of
Section~\ref{sec-user}. Each entry represents a rubric item, and for
each item it specifies (i) what the item applies to (an assignment
component/question, all questions, or the assignment overall); (ii) the
total number of points for the component/question (only for items
specific to a component/question); (iii) the prompt code the grader
needs to type to apply the item; (iv) the prompt message to be shown
next to the prompt code; (v) the feedback to apply when the item is
selected; and (vi) the points to remove from (or to add to) the total
grade when the item is selected. The function
\texttt{create\_rubric\_template()} creates an empty csv file with the
necessary column names to aid in formatting the rubric.

The meaning of each column of the rubric formatted for gradetools is
further illustrated in Table~\ref{tbl-rubric}, with an example rubric
item applicable to all questions (note that total points are left blank
for items applicable to all questions, as shown in
Figure~\ref{fig-grading-example-formatted-rubric}). Let's assume that
the grader wants to provide the feedback ``Please adhere to the
Tidyverse style guide'' for a specific question or for the overall
assignment. This is the feedback that the student will see. However,
writing out this comment each time the grader encounters a submission
that needs such feedback is time consuming. To save time, gradetools
utilizes a short code, that we denote prompt code, in this case defined
as 1. By selecting 1, the grader is able to provide the full length
comment and apply the corresponding score policy. To be quick-to-enter,
prompt codes can be short and uninformative and therefore difficult to
remember, so a short prompt message can be shown along with the prompt
code to remind the grader what feedback the prompt code corresponds to.

\hypertarget{tbl-rubric}{}
\begin{table}
\caption{\label{tbl-rubric}The specifics of each entry (item) in the rubric, in the case of an item
applicable to all questions. }\tabularnewline

\centering
\fontsize{11}{13}\selectfont
\begin{tabular}{>{\raggedright\arraybackslash}p{1.5in}>{\raggedright\arraybackslash}p{2.2in}>{\raggedright\arraybackslash}p{2.2in}}
\toprule
Specifics & Description & Example\\
\midrule
\cellcolor{gray!6}{Name} & \cellcolor{gray!6}{What question the item is available for} & \cellcolor{gray!6}{All questions}\\
Prompt code & What the grader enters to apply this item & 1\\
\cellcolor{gray!6}{Prompt message} & \cellcolor{gray!6}{The description that the grader sees while grading} & \cellcolor{gray!6}{tidyverse code style}\\
Feedback & What the student sees when they receive feedback & Please adhere to the Tidyverse style guide, as discussed in Lecture 1.\\
\cellcolor{gray!6}{Points to remove} & \cellcolor{gray!6}{The penalty applied when this item is selected} & \cellcolor{gray!6}{0.5 points}\\
\bottomrule
\end{tabular}
\end{table}

In gradetools, while grading, the grader is able to see the prompt
message and the prompt code, while the student will be able to see the
corresponding personalized and extended feedback when feedback files are
returned. As shown in Figure~\ref{fig-grading-example-student2}, the
grader can quickly indicate (possibly multiple) items to apply among
available ones in the rubric.

As anticipated in Section~\ref{sec-concept}, linking each rubric item
with some corresponding feedback allows gradetools to scale the
provision of detailed individualized feedback. In addition to expediting
provision of repeatable feedback across different questions and
students, gradetools supports the provision of unique feedback on the
fly and the dynamic update of the rubric, for which prompt codes ``p''
and ``r'' are reserved, respectively (as shown in
Figure~\ref{fig-grading-example-student1}).

\hypertarget{sec-maintain-grade}{%
\subsection{Maintaining the grade sheet and the feedback
files}\label{sec-maintain-grade}}

\begin{figure}

{\centering \includegraphics[width=145mm,height=\textheight]{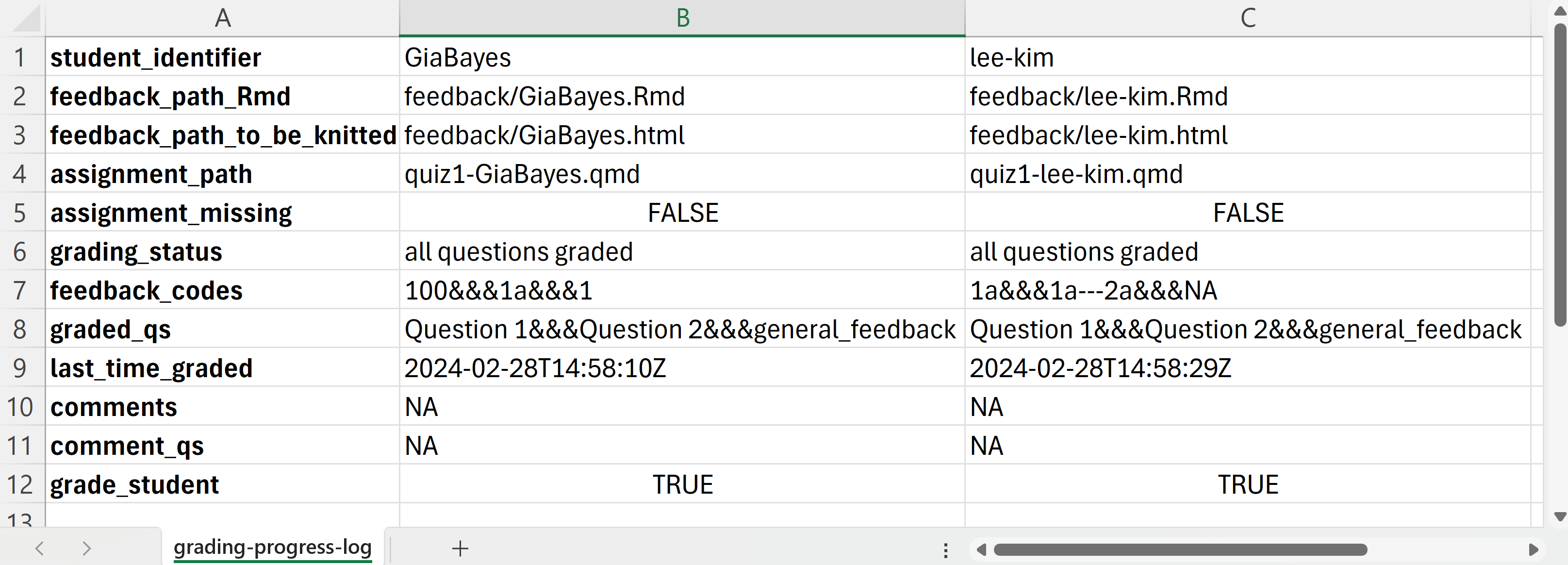}

}

\caption{\label{fig-example-grading-log}Screenshot of grading progress
log file, at the end of grading two students as in the example from
Section~\ref{sec-grading-ex}.}

\end{figure}

Finally, we consider how gradetools assists with maintaining the grade
sheet and feedback files. Grading progress is recorded by gradetools in
a \emph{grading progress log file}. Figure~\ref{fig-example-grading-log}
shows the content of this file for the example of
Section~\ref{sec-grading-ex}. When beginning to grade a new assignment,
this file is created with a row for each student in the provided roster
(displayed as column in Figure~\ref{fig-example-grading-log}, for
readability), including information such as where assignments files are
located and indicating that all students are ungraded. When grading a
student's submission, the grading log file is dynamically updated to
keep track of any prompt code entered by the grader and eventually
students grading statuses change from ``ungraded'' to ``all questions
graded''. The grading progress log file is not meant to be edited
directly by the grader, but is used by gradetools to keep track of the
grading progress so that grading can be interrupted and resumed at any
time. Specifically, as shown in Figure~\ref{fig-example-grading-log},
the grading progress log saves all rubric prompt codes that have been
applied by the grader. At the end of the grading process, as discussed
in Section~\ref{sec-assign-grade}, applied prompt codes are used in
combination with the rubric to produce the grades for the final grade
sheet and the feedback for each student's feedback file, e.g., the ones
shown in Figure~\ref{fig-grading-example-outputs}. Only storing the
selected prompt codes allows the grader to change the grade and feedback
corresponding to any rubric item, anytime until grading is completed,
and see the changes be reflected in the feedback files and final grade
sheet (as mentioned in Section~\ref{sec-grading-ex}). Importantly, a
grader should never change prompt codes that have already been used, as
this would result in the loss of grading progress.

At the end of grading, a feedback file is created for each student. When
calling gradetools' functions, the grader specifies where the feedback
files should be stored by providing an example feedback path. Similarly
as with the example assignment path described in
Section~\ref{sec-open-submis}, the provided student identifier must be
present in the provided example feedback path, so that files with
feedback for different submissions can always be distinguished.
Information on where the grader wishes to store feedback files is also
recorded in the grading progress log, as shown in
Figure~\ref{fig-example-grading-log}.

Since there are a plethora of ways instructors distribute grades and
feedback, automated delivery of grades and feedback files to the
students was determined to be outside of gradetools' scope, with the
exception of pushing feedback files to GitHub. In the next section, we
provide some alternatives in the R ecosystem.

\hypertarget{sec-discussion}{%
\section{Discussion}\label{sec-discussion}}

In Section~\ref{sec-concept} we have discussed the grading workflow as a
three-step process: preparation, grading and providing feedback, and
finalization. In Section~\ref{sec-user} we have shown what automation of
this grading workflow looks like with the R package gradetools and in
Section~\ref{sec-developer} we have detailed how gradetools supports
this automation. Figure~\ref{fig-gradetools-diagram} summarizes
gradetools functions, emphasizing the pedagogical tasks, and noting the
tasks managed by each function. From this figure, it is evident that
gradetools has some limitations and does not serve every step of the
grading process.

In terms of inputs, graders will need to provide the roster and
students' submissions and gradetools does not support retrieval of them.
In terms of outputs, the package provides feedback files for each
student and the overall gradesheet. Gradetools only supports returning
of feedback files to students via GitHub. For the tasks that are not
supported by gradetools, there are many R packages that can facilitate
these tasks such as \textbf{rcanvas} \citep{R-rcanvas}, \textbf{moodleR}
\citep{R-moodleR}, \textbf{ghclass} \citep{R-ghclass}, \textbf{gmailr}
\citep{R-gmailr} for working with the Canvas, Moodle, GitHub, and GMail
Application Programming Interfaces (APIs) respectively. These packages
can support tasks like retrieving students' work and returning students'
scores and feedback. Some of the Learning Management Systems (LMSs) may
also provide interfaces that allow bulk downloads and uploads manually.

The gradetools package focuses mainly on the second step of the grading
workflow by improving the grading and feedback process through
automating the administrative tasks. The most important benefit of using
gradetools is that it helps adopt an efficient and fair grading
workflow. Even though we did not study it rigorously, in terms of
efficiency, gradetools saves a lot of time in grading once the initial
learning curve has been passed. In terms of fairness, the fact that
gradetools enforces use of a rubric allows for consistent grading and
feedback across different students and questions. Use of rubrics are
pivotal to fairness especially in performance-based assessments
\citep{shepherd2008rubrics}.

In summary, gradetools automates many administrative tasks in the
grading workflow with many pedagogical considerations but it is by no
means a single solution to a fully automated grading workflow.
Instructors who are interested in fully automating the grading workflow,
would need to be proficient in R and rely on packages other than
gradetools. For instance, if an instructor downloads files from an LMS
they might need to do string manipulation to have filing name
consistency across different students' file names. In our courses, we
have managed to fully automate our grading workflow by supplementing
gradetools with GitHub features of ghclass \citep{R-ghclass} and data
wrangling features of the tidyverse packages \citep{tidyverse}.

The gradetools package can be an important addition to an instructor's
toolkit, especially to support reproducibility. With a stronger emphasis
on teaching of reproducibility skills, students working directly in
literate programming notebooks such as Quarto and being able to manage
and name these files is an important part of their data science training
\citep{pruim2023fostering}. The gradetools package is designed to
support such teaching scenarios. In addition to teaching
reproducibility, reproducible teaching is also important
\citep{dogucu2022tools}. The gradetools package allows for grading code
and files (e.g., rubric) to be reused (or modified) from one academic
term to the next, and to be picked up by other instructors or TAs
grading the same assignment.

In addition to increased emphasis on reproducibility, another important
change in data science education that can motivate the use of
automated-grading-workflow tools is the development of generative
artifical intelligence (AI) tools. Even though there is not a consensus
on how AI tools can benefit the learning process, examples of use show
that AI is here to stay to be part of the learning process
\citep{ellis}. With students' increased use of AI, we expect instructors
to modify some of their assignments to alternative formats with more
reliance on open-ended assessments, deriving a greater need for tools
like gradetools.

Contrary to some other grading tools, since gradetools is an R package,
it is free to use. It does not require internet connection during the
extensive period of grading and providing feedback. However, users would
still need internet connection in the first and third steps of the
grading workflow.

When students' work is considered, needless to say privacy is important
(e.g., grades) and can be protected under law depending on the country.
In our grading process we have used the package on our local computers
and stored the grade sheets and other private documents locally.
However, it is worth noting that users who choose to use R and
gradetools on different platforms such as in the Cloud will need to be
mindful of what they are storing, what is legal and ethical to store in
that specific platform.

In this paper, in addition to introducing gradetools and how it can be
utilized in data science classes, we have also shared our vision of an
automated grading workflow and defined the distinction between
pedagogical tasks and administrative tasks in grading, defined different
feedback types such as unique and repeated ones. We will continue to
maintain gradetools for use in our own data science courses and beyond.
We hope that this work will help the community of data science and
statistics educators use gradetools as their grading workflow assistant
or develop their own tools for assisting their grading workflow.

  \bibliography{references.bib}

\end{document}